%Paper: hep-th/9209121
%From: kleppe%vtinte.dnet@vtserf.cc.vt.edu
%Date: Tue, 29 Sep 92 11:51:10 -0400
%Date (revised): Tue, 29 Sep 92 13:32:34 -0400

\catcode`@=11
\expandafter\ifx\csname inp@t\endcsname\relax\let\inp@t=\input
\def\input#1 {\expandafter\ifx\csname #1IsLoaded\endcsname\relax
\inp@t#1%
\expandafter\def\csname #1IsLoaded\endcsname{(#1 was previously loaded)}
\else\message{\csname #1IsLoaded\endcsname}\fi}\fi
\catcode`@=12

\font\twelverm=cmr12			\font\twelvei=cmmi12
\font\twelvesy=cmsy10 scaled 1200	\font\twelveex=cmex10 scaled 1200
\font\twelvebf=cmbx12			\font\twelvesl=cmsl12
\font\twelvett=cmtt12			\font\twelveit=cmti12
\font\twelvesc=cmcsc10 scaled 1200	\font\twelvesf=cmss12
                     
\font\twelvemib=cmmib10 scaled 1200
\font\tenmib=cmmib10
\font\eightmib=cmmib10 scaled 800

\skewchar\twelvei='177			\skewchar\twelvesy='60
\skewchar\twelvemib='177
\newfam\mibfam
\def\twelvepoint{\normalbaselineskip=12.4pt plus 0.1pt minus 0.1pt
  \abovedisplayskip 12.4pt plus 3pt minus 9pt
  \belowdisplayskip 12.4pt plus 3pt minus 9pt
  \abovedisplayshortskip 0pt plus 3pt
  \belowdisplayshortskip 7.2pt plus 3pt minus 4pt
  \smallskipamount=3.6pt plus1.2pt minus1.2pt
  \medskipamount=7.2pt plus2.4pt minus2.4pt
  \bigskipamount=14.4pt plus4.8pt minus4.8pt
  \def\rm{\fam0\twelverm}          \def\it{\fam\itfam\twelveit}%
  \def\sl{\fam\slfam\twelvesl}     \def\bf{\fam\bffam\twelvebf}%
  \def\mit{\fam 1}                 \def\cal{\fam 2}%
  \def\sc{\twelvesc}		   \def\tt{\twelvett}%
  \def\sf{\twelvesf}               \def\mib{\fam\mibfam\twelvemib}%
  \textfont0=\twelverm   \scriptfont0=\tenrm   \scriptscriptfont0=\sevenrm
  \textfont1=\twelvei    \scriptfont1=\teni    \scriptscriptfont1=\seveni
  \textfont2=\twelvesy   \scriptfont2=\tensy   \scriptscriptfont2=\sevensy
  \textfont3=\twelveex   \scriptfont3=\twelveex\scriptscriptfont3=\twelveex
  \textfont\itfam=\twelveit
  \textfont\slfam=\twelvesl
  \textfont\bffam=\twelvebf \scriptfont\bffam=\tenbf
                            \scriptscriptfont\bffam=\sevenbf
  \textfont\mibfam=\twelvemib \scriptfont\mibfam=\tenmib
                              \scriptscriptfont\mibfam=\eightmib
  \normalbaselines\rm}

\mathchardef\alpha="710B
\mathchardef\beta="710C
\mathchardef\gamma="710D
\mathchardef\delta="710E
\mathchardef\epsilon="710F
\mathchardef\zeta="7110
\mathchardef\eta="7111
\mathchardef\theta="7112
\mathchardef\iota="7113
\mathchardef\kappa="7114
\mathchardef\lambda="7115
\mathchardef\mu="7116
\mathchardef\nu="7117
\mathchardef\xi="7118
\mathchardef\pi="7119
\mathchardef\rho="711A
\mathchardef\sigma="711B
\mathchardef\tau="711C
\mathchardef\phi="711E
\mathchardef\chi="711F
\mathchardef\psi="7120
\mathchardef\omega="7121
\mathchardef\varepsilon="7122
\mathchardef\vartheta="7123
\mathchardef\varpi="7124
\mathchardef\varrho="7125
\mathchardef\varsigma="7126
\mathchardef\varphi="7127
\def\beginlinemode{\endmode
  \begingroup\parskip=0pt \obeylines\def\\{\par}\def\endmode{\par\endgroup}}
\def\beginparmode{\endmode
  \begingroup \def\endmode{\par\endgroup}}
\let\endmode=\par
{\obeylines\gdef\
{}}
\def\singlespace{\baselineskip=\normalbaselineskip}

\def\oneandahalfspace{\baselineskip=\normalbaselineskip
  \multiply\baselineskip by 3 \divide\baselineskip by 2}
\def\doublespace{\baselineskip=\normalbaselineskip \multiply\baselineskip by 2}

\newcount\firstpageno
\firstpageno=2
\footline={\ifnum\pageno<\firstpageno{\hfil}\else{\hfil\twelverm\folio\hfil}\fi}
\def\toppageno{\global\footline={\hfil}\global\headline
  ={\ifnum\pageno<\firstpageno{\hfil}\else{\hfil\twelverm\folio\hfil}\fi}}
\let\rawfootnote=\footnote		% We must set the footnote style
\def\footnote#1#2{{\rm\singlespace\parindent=0pt\parskip=0pt
  \rawfootnote{#1}{#2\hfill\vrule height 0pt depth 6pt width 0pt}}}
\def\raggedcenter{\leftskip=4em plus 12em \rightskip=\leftskip
  \parindent=0pt \parfillskip=0pt \spaceskip=.3333em \xspaceskip=.5em
  \pretolerance=9999 \tolerance=9999
  \hyphenpenalty=9999 \exhyphenpenalty=9999 }
\def\dateline{\rightline{\ifcase\month\or
  January\or February\or March\or April\or May\or June\or
  July\or August\or September\or October\or November\or December\fi
  \space\number\year}}
\def\received{\vskip 3pt plus 0.2fill
 \centerline{\sl (Received\space\ifcase\month\or
  January\or February\or March\or April\or May\or June\or
  July\or August\or September\or October\or November\or December\fi
  \qquad, \number\year)}}
\hsize=6.5truein
\vsize=8.9truein
\def\nooffset{\global\hoffset=0pt\global\voffset=0pt}

\nooffset
\parskip=\medskipamount
\def\\{\cr}
\twelvepoint		% selects twelvepoint fonts (cf. \tenpoint)
\doublespace		% selects double spacing for main part of paper (cf.
			%	\singlespace, \oneandahalfspace)
\overfullrule=0pt	% delete the nasty little black boxes for overfull box
  %  "Draft", Timestamp

\def\preprintno#1{
 \rightline{\rm #1}}	% Preprint number at upper right of title page

\def\title			%  Title on title page
  {\null\vskip 3pt plus 0.2fill
   \beginlinemode \doublespace \raggedcenter \bf}

\def\author			%  Author(s) name(s)  on title page
  {\vskip 3pt plus 0.2fill \beginlinemode
   \singlespace \raggedcenter\rm}

\def\affil			% Affiliations (can intermix with \author)
  {\beginlinemode\oneandahalfspace \raggedcenter \sl}

\def\abstract			% Begin abstract
  {\vskip 3pt plus 0.3fill \beginparmode
   \oneandahalfspace ABSTRACT: }

\def\endtitlepage		% End title page, begin body of paper
  {\endpage			% 	This subsumes \body
   \body}
\let\endtopmatter=\endtitlepage

\def\body			% Begin text body;  can be used to end
  {\beginparmode}		% \title, \author, \affil, \abstract,
				% \reference, or \figurecaption modes

\def\head#1{			% Head;  NOTE enclose the text in {}
  \goodbreak\vskip 0.4truein	%  e.g., \head{I. Introduction}
  {\immediate\write16{#1}
   \raggedcenter {\sc #1}\par}
   \nobreak\vskip 0truein\nobreak}

\def\itemitemitem{\par\indent\indent \hangindent3\parindent \textindent}
\def\itemitemitemitem{\par\indent\indent\indent \hangindent4\parindent
\textindent}
\def\beginitems{\par\medskip\bgroup
  \def\i##1 {\par\noindent\llap{##1\enspace}\ignorespaces}%
  \def\ii##1 {\item{##1}}%
  \def\iii##1 {\itemitem{##1}}%
  \def\iiii##1 {\itemitemitem{##1}}%
  \def\iiiii##1 {\itemitemitemitem{##1}}
  \leftskip=36pt\parskip=0pt}\def\enditems{\par\egroup}

\def\makefigure#1{\parindent=36pt\item{}Figure #1}

\def\figure#1 (#2) #3\par{\goodbreak\midinsert
\vskip#2
\bgroup\makefigure{#1} #3\par\egroup\endinsert}

\def\beneathrel#1\under#2{\mathrel{\mathop{#2}\limits_{#1}}}

\def\refto#1{$^{#1}$}		% For references in text as superscript

\def\references			% Begin references -- basic format is Phys Rev
  {\head{References}		% I.e., volume, page, year (space after commas).
   \beginparmode
   \frenchspacing \parindent=0pt \leftskip=1truecm
   \parskip=8pt plus 3pt \everypar{\hangindent=\parindent}}

\gdef\refis#1{\item{#1.\ }}			% Ref list numbers.

\gdef\journal#1, #2, #3, 1#4#5#6{		% Journal reference.  Comma sets
    {\sl #1~}{\bf #2}, #3 (1#4#5#6)}		% off: name, vol, page, year

\def\refstylenp{		% Nucl Phys(or Phys Lett) ref style: V, Y, P
  \gdef\refto##1{ [##1]}			% Reference in text []
  \gdef\refis##1{\item{##1)\ }}			% Reference list numbers)
  \gdef\journal##1, ##2, ##3, ##4 {		% Journal reference
     {\sl ##1~}{\bf ##2~}(##3) ##4 }}

\def\rmp{\journal Rev. Mod. Phys., }

\def\endreferences{\body}

\def\figurecaptions		% Begin figure captions
  {\endpage
   \beginparmode
   \head{Figure Captions}
}

\def\endpage			%  Eject a page
  {\vfill\eject}

\def\endpaper			%  Ways to say goodbye
  {\endmode\vfill\supereject}

\def\heading				% Heading
  {\vskip 0.5truein plus 0.1truein	% e.g., \heading I. NOTES \endheading
   \beginparmode \def\\{\par} \parskip=0pt \singlespace \raggedcenter}

\def\subheading				% Subheading
  {\vskip 0.25truein plus 0.1truein	% e.g., \subheading{A. The Problem}
   \beginlinemode \singlespace \parskip=0pt \def\\{\par}\raggedcenter}

\def\tag#1$${\eqno(#1)$$}

\def\align#1$${\eqalign{#1}$$}

\def\aligntag#1$${\gdef\tag##1\\{&(##1)\cr}\eqalignno{#1\\}$$
  \gdef\tag##1$${\eqno(##1)$$}}

\def\endaligntag{}

\def\overset #1\to#2{{\mathop{#2}\limits^{#1}}}
\def\underset#1\to#2{{\let\next=#1\mathpalette\undersetpalette#2}}
\def\undersetpalette#1#2{\vtop{\baselineskip0pt
\ialign{$\mathsurround=0pt #1\hfil##\hfil$\crcr#2\crcr\next\crcr}}}

\def\ref#1{Ref.~#1}			% 	for inline references
\def\Ref#1{Ref.~#1}			% 	ditto
\def\[#1]{[\cite{#1}]}
\def\cite#1{{#1}}
			% For figure numbers
		% For citation of equation numbers
	%	ditto
			%	ditto
			%	ditto
		%	ditto
\def\(#1){(\call{#1})}
\def\call#1{{#1}}
\def\taghead#1{}
\def\frac#1#2{{\textstyle {#1 \over #2}}}
\def\half{{\frac 12}}

\def\12{{1\over2}}

\def\sla{\raise.15ex\hbox{$/$}\kern-.57em}
\def\leaderfill{\leaders\hbox to 1em{\hss.\hss}\hfill}
\def\twiddle{\lower.9ex\rlap{$\kern-.1em\scriptstyle\sim$}}
\def\bigtwiddle{\lower1.ex\rlap{$\sim$}}
\def\gtwid{\mathrel{\raise.3ex\hbox{$>$\kern-.75em\lower1ex\hbox{$\sim$}}}}
\def\ltwid{\mathrel{\raise.3ex\hbox{$<$\kern-.75em\lower1ex\hbox{$\sim$}}}}
\def\square{\kern1pt\vbox{\hrule height 1.2pt\hbox{\vrule width 1.2pt\hskip 3pt
   \vbox{\vskip 6pt}\hskip 3pt\vrule width 0.6pt}\hrule height 0.6pt}\kern1pt}
\def\tdot#1{\mathord{\mathop{#1}\limits^{\kern2pt\ldots}}}
\def\happyface{%
$\bigcirc\rlap{\lower0.3ex\hbox{$\kern-0.85em\scriptscriptstyle\smile$}%
\raise0.4ex\hbox{$\kern-0.6em\scriptstyle\cdot\cdot$}}$}
\def\sadface{%
$\bigcirc\rlap{\lower0.25ex\hbox{$\kern-0.85em\scriptscriptstyle\frown$}%
\raise0.43ex\hbox{$\kern-0.6em\scriptstyle\cdot\cdot$}}$}

\def\pmb#1{\setbox0=\hbox{#1}%
  \kern-.025em\copy0\kern-\wd0
  \kern  .05em\copy0\kern-\wd0
  \kern-.025em\raise.0433em\box0 }

\def\exp{{\rm exp}}

\def\vt{Department of Physics\\Virginia Polytechnic Institute\\and State %
University\\Blacksburg VA 24060}

	% Sorry, no def. for \coffeetable

\def\m@th{\mathsurround=0pt}
\def\leftrightarrowfill{$\m@th \mathord\leftarrow \mkern-6mu
 \cleaders\hbox{$\mkern-2mu \mathord- \mkern-2mu$}\hfill
 \mkern-6mu \mathord\rightarrow$}
\def\overleftrightarrow#1{\vbox{ialign{##\crcr
	\leftrightarrowfill\crcr\noalign{\kern-1pt\nointerlineskip}
	$\hfil\displaystyle{#1}\hfil$\crcr}}}
\catcode`@=11
\newcount\tagnumber\tagnumber=0
\immediate\newwrite\eqnfile
\newif\if@qnfile\@qnfilefalse
\def\write@qn#1{}
\def\writenew@qn#1{}
\def\w@rnwrite#1{\write@qn{#1}\message{#1}}
\def\@rrwrite#1{\write@qn{#1}\errmessage{#1}}

\def\taghead#1{\gdef\t@ghead{#1}\global\tagnumber=0}
\def\t@ghead{}

\expandafter\def\csname @qnnum-3\endcsname
  {{\t@ghead\advance\tagnumber by -3\relax\number\tagnumber}}
\expandafter\def\csname @qnnum-2\endcsname
  {{\t@ghead\advance\tagnumber by -2\relax\number\tagnumber}}
\expandafter\def\csname @qnnum-1\endcsname
  {{\t@ghead\advance\tagnumber by -1\relax\number\tagnumber}}
\expandafter\def\csname @qnnum0\endcsname
  {\t@ghead\number\tagnumber}
\expandafter\def\csname @qnnum+1\endcsname
  {{\t@ghead\advance\tagnumber by 1\relax\number\tagnumber}}
\expandafter\def\csname @qnnum+2\endcsname
  {{\t@ghead\advance\tagnumber by 2\relax\number\tagnumber}}
\expandafter\def\csname @qnnum+3\endcsname
  {{\t@ghead\advance\tagnumber by 3\relax\number\tagnumber}}

\def\equationfile{%
  \@qnfiletrue\immediate\openout\eqnfile=\jobname.eqn%
  \def\write@qn##1{\if@qnfile\immediate\write\eqnfile{##1}\fi}
  \def\writenew@qn##1{\if@qnfile\immediate\write\eqnfile
    {\noexpand\tag{##1} = (\t@ghead\number\tagnumber)}\fi}
}

\def\callall#1{\xdef#1##1{#1{\noexpand\call{##1}}}}
\def\call#1{\each@rg\callr@nge{#1}}

\def\each@rg#1#2{{\let\thecsname=#1\expandafter\first@rg#2,\end,}}
\def\first@rg#1,{\thecsname{#1}\apply@rg}
\def\apply@rg#1,{\ifx\end#1\let\next=\relax%
\else,\thecsname{#1}\let\next=\apply@rg\fi\next}

\def\callr@nge#1{\calldor@nge#1-\end-}
\def\callr@ngeat#1\end-{#1}
\def\calldor@nge#1-#2-{\ifx\end#2\@qneatspace#1 %
  \else\calll@@p{#1}{#2}\callr@ngeat\fi}
\def\calll@@p#1#2{\ifnum#1>#2{\@rrwrite{Equation range #1-#2\space is bad.}
\errhelp{If you call a series of equations by the notation M-N, then M and
N must be integers, and N must be greater than or equal to M.}}\else%
 {\count0=#1\count1=#2\advance\count1
by1\relax\expandafter\@qncall\the\count0,%
  \loop\advance\count0 by1\relax%
    \ifnum\count0<\count1,\expandafter\@qncall\the\count0,%
  \repeat}\fi}

\def\@qneatspace#1#2 {\@qncall#1#2,}
\def\@qncall#1,{\ifunc@lled{#1}{\def\next{#1}\ifx\next\empty\else
  \w@rnwrite{Equation number \noexpand\(>>#1<<) has not been defined yet.}
  >>#1<<\fi}\else\csname @qnnum#1\endcsname\fi}

\let\eqnono=\eqno
\def\eqno(#1){\tag#1}
\def\tag#1$${\eqnono(\displayt@g#1 )$$}

\def\aligntag#1\endaligntag
  $${\gdef\tag##1\\{&(##1 )\cr}\eqalignno{#1\\}$$
  \gdef\tag##1$${\eqnono(\displayt@g##1 )$$}}

\def\eqalignno#1{\displ@y \tabskip\centering
  \halign to\displaywidth{\hfil$\displaystyle{##}$\tabskip\z@skip
    &$\displaystyle{{}##}$\hfil\tabskip\centering
    &\llap{$\displayt@gpar##$}\tabskip\z@skip\crcr
    #1\crcr}}

\def\displayt@gpar(#1){(\displayt@g#1 )}

\def\displayt@g#1 {\rm\ifunc@lled{#1}\global\advance\tagnumber by1
        {\def\next{#1}\ifx\next\empty\else\expandafter
        \xdef\csname @qnnum#1\endcsname{\t@ghead\number\tagnumber}\fi}%
  \writenew@qn{#1}\t@ghead\number\tagnumber\else
        {\edef\next{\t@ghead\number\tagnumber}%
        \expandafter\ifx\csname @qnnum#1\endcsname\next\else
        \w@rnwrite{Equation \noexpand\tag{#1} is a duplicate number.}\fi}%
  \csname @qnnum#1\endcsname\fi}

\def\ifunc@lled#1{\expandafter\ifx\csname @qnnum#1\endcsname\relax}

\let\@qnend=\end\gdef\end{\if@qnfile
\immediate\write16{Equation numbers written on []\jobname.EQN.}\fi\@qnend}

\catcode`@=12
\refstylenp
\catcode`@=11
\newcount\r@fcount \r@fcount=0
\def\refreset{\global\r@fcount=0}
\newcount\r@fcurr
\immediate\newwrite\reffile
\newif\ifr@ffile\r@ffilefalse
\def\w@rnwrite#1{\ifr@ffile\immediate\write\reffile{#1}\fi\message{#1}}

\def\writer@f#1>>{}
\def\referencefile{%			  Stuff to write .REF file
  \r@ffiletrue\immediate\openout\reffile=\jobname.ref%
  \def\writer@f##1>>{\ifr@ffile\immediate\write\reffile%
    {\noexpand\refis{##1} = \csname r@fnum##1\endcsname = %
     \expandafter\expandafter\expandafter\strip@t\expandafter%
     \meaning\csname r@ftext\csname r@fnum##1\endcsname\endcsname}\fi}%
  \def\strip@t##1>>{}}

\def\citeall#1{\xdef#1##1{#1{\noexpand\cite{##1}}}}
\def\cite#1{\each@rg\citer@nge{#1}}	% Variable No. of args, separated by ","

\def\each@rg#1#2{{\let\thecsname=#1\expandafter\first@rg#2,\end,}}
\def\first@rg#1,{\thecsname{#1}\apply@rg}	% each@ag is a general purpose
\def\apply@rg#1,{\ifx\end#1\let\next=\relax%	  variable no. of arg. macro.
\else,\thecsname{#1}\let\next=\apply@rg\fi\next}% args separated by commas

\def\citer@nge#1{\citedor@nge#1-\end-}	% Check for M-N range (M and N numbers)
\def\citer@ngeat#1\end-{#1}
\def\citedor@nge#1-#2-{\ifx\end#2\r@featspace#1 % Single argument
  \else\citel@@p{#1}{#2}\citer@ngeat\fi}	% M-N range of arguments
\def\citel@@p#1#2{\ifnum#1>#2{\errmessage{Reference range #1-#2\space is bad.}%
    \errhelp{If you cite a series of references by the notation M-N, then M and
    N must be integers, and N must be greater than or equal to M.}}\else%
 {\count0=#1\count1=#2\advance\count1
by1\relax\expandafter\r@fcite\the\count0,%
  \loop\advance\count0 by1\relax%	  Loop from M to N
    \ifnum\count0<\count1,\expandafter\r@fcite\the\count0,%
  \repeat}\fi}

\def\r@featspace#1#2 {\r@fcite#1#2,}	% Eat spaces at beginning or end of arg
\def\r@fcite#1,{\ifuncit@d{#1}%		  Cite individual reference
    \newr@f{#1}%
    \expandafter\gdef\csname r@ftext\number\r@fcount\endcsname%
                     {\message{Reference #1 to be supplied.}%
                      \writer@f#1>>#1 to be supplied.\par}%
 \fi%
 \csname r@fnum#1\endcsname}
\def\ifuncit@d#1{\expandafter\ifx\csname r@fnum#1\endcsname\relax}%
\def\newr@f#1{\global\advance\r@fcount by1%
    \expandafter\xdef\csname r@fnum#1\endcsname{\number\r@fcount}}

\let\r@fis=\refis			% Save old \refis, redefine
\def\refis#1#2#3\par{\ifuncit@d{#1}%      Use two params #2 #3 to strip blank
   \newr@f{#1}%
   \w@rnwrite{Reference #1=\number\r@fcount\space is not cited up to now.}\fi%
  \expandafter\gdef\csname r@ftext\csname r@fnum#1\endcsname\endcsname%
  {\writer@f#1>>#2#3\par}}

\def\ignoreuncited{%   redefine \refis if ignoring uncited references
   \def\refis##1##2##3\par{\ifuncit@d{##1}%
     \else\expandafter\gdef\csname r@ftext\csname
r@fnum##1\endcsname\endcsname%
     {\writer@f##1>>##2##3\par}\fi}}

\def\r@ferr{\endreferences\errmessage{I was expecting to see
\noexpand\endreferences before now;  I have inserted it here.}}
\let\r@ferences=\references
\def\references{\r@ferences\def\endmode{\r@ferr\par\endgroup}}

\let\endr@ferences=\endreferences
\def\endreferences{\r@fcurr=0%		  Save old \endreferences, redefine
  {\loop\ifnum\r@fcurr<\r@fcount%	  Loop over refnum and produce text
    \advance\r@fcurr by 1\relax\expandafter\r@fis\expandafter{\number\r@fcurr}%
    \csname r@ftext\number\r@fcurr\endcsname%
  \repeat}\gdef\r@ferr{}\global\r@fcount=0\endr@ferences}

% Save old \endpaper, redefine it to write parting message.

\let\r@fend=\endpaper\gdef\endpaper{\ifr@ffile
\immediate\write16{Cross References written on []\jobname.REF.}\fi\r@fend}

\catcode`@=12

\citeall\refto		% These macros will generate citations
\citeall\ref		%
\citeall\Ref		%
\nooffset
\preprintno{VPI-IHEP-92/12}
\title{Transformation Properties of Linearized de Sitter Gravity Solutions}
\vskip 0.2in
\author{Gary Kleppe}
\vskip 0.2in
\affil\vt
KLEPPE @ VTINTE . BITNET
KLEPPE\%VTINTE.DNET@VTSERF.CC.VT.EDU
\abstract{The effect of de Sitter transformations on Tsamis and Woodard's
solutions to the linearized gauge fixed equations of motion of quantum gravity
in a de Sitter space background is worked out explicitly. It is shown that
these solutions
are closed under the transformations of the de Sitter group. To do this it is
necessary to use a compensating gauge transformation to return the transformed
solution to the original gauge.}
\endtopmatter
\nooffset

De Sitter space is the unique maximally symmetric solution to Einstein's
gravitational equations with a positive cosmological constant. Therefore,
it is interesting to study quantum effects on de Sitter space, as they may
provide a solution\refto{strong} to the cosmological constant
problem\refto{weinberg}. Of course perturbative quantum gravity is
nonrenormalizable, but it can still be used as a low-energy effective
theory to study infrared effects.

Recently the propagator for quantum gravity in a de Sitter space background
has been found\refto{structure}, and the linearized gauge-fixed equations of
motion solved\refto{mode}. This has been accomplished by using a conformal
coordinate system which only covers half of the full de Sitter space manifold.
However, the de Sitter group, the fundamental set of symmetries of the de
Sitter
background, contains transformations which take us out of this restricted
submanifold. For this reason, and also because the gauge which was used is not
de Sitter invariant, it is not obvious that the solutions of the equations
of motion will transform into themselves under the de Sitter group.

In this paper we will show that these gauge-fixed solutions still possess
de Sitter invariance. We will show that since the gauge used is not invariant,
it is necessary to combine de Sitter transformations with gauge transformations
which return the de Sitter transformed solutions to the chosen gauge. We work
out in detail the de Sitter transformation of an arbitrary solution, and show
explicitly that the set of solutions is closed under the de Sitter
transformations.

$D$ dimensional de Sitter space may be defined as the surface of constant
length $1/H$ from the origin of $D+1$ dimensional Minkowski space. The $D$
coordinates $x^0,\ x^i$ necessary to describe de Sitter space may then be
constructed as functions of the Minkowski embedding space coordinates
$X^0,\ X^i,\ X^D$. We will use the conformal coordinate system of
Tsamis and Woodard\refto{structure}:
$$u\equiv x^0={{\sqrt{X\cdot X}}\over H(X^0+X^D)}\eqno(imbed a)$$
$$x^i={X^i\over H(X^0+X^D)}\eqno(imbed b)$$
The inverse of this mapping is
$$X^0={1+H^2(x^ix^i-u^2)\over 2H^2u}\eqno(outofbed a)$$
$$X^i={x^i\over Hu}\eqno(outofbed b)$$
$$X^D={1-H^2(x^ix^i-u^2)\over 2H^2u}\eqno(outofbed c)$$
The induced de Sitter space metric is then
$$ds^2={-du^2+dx^idx^i\over H^2u^2},\eqno(metric)$$
hence the name ``conformal''.

The de Sitter group is the Lorentz group of the $D+1$ dimensional embedding
space. It will be convenient to parametrize it as follows:
$$\delta X^0=\half (d^i-a^i) X^i+bX^D\eqno(Xtrans a)$$
$$\delta X^i=\half d^i (X^0+X^D)-\half a^i (X^0-X^D)+c^{ij}X^j\eqno(Xtrans b)$$
$$\delta X^D=bX^0-\half(a^i+d^i) X^i\eqno(Xtrans c)$$
Using \(imbed), it is straightforward to calculate the effect of these
transformations on the de Sitter coordinates:
$$\delta u =-bu+Ha^iux^i\eqno(xtrans a)$$
$$\delta x^i=\frac1{2H}d^i+\half Ha^i(u^2-x^jx^j)+Ha^jx^ix^j+c^{ij}x^j-bx^i
\eqno(xtrans b)$$
Note that these rules may be written in a pseudo-covariant form
$$\delta x^\mu=\frac1{2H}d^\mu+c^{\mu\nu}-bx^\mu-\half Ha^\mu x^\nu x_\nu
+Ha^\nu x^\mu x_\nu\eqno()$$
if we take all zero components of $a,\ c,$ and $d$ to be zero.

Now we allow the metric to vary about the de Sitter space background:
$$g_{\mu\nu}={\hat g}_{\mu\nu}+\kappa h_{\mu\nu}\eqno()$$
where
$${\hat g}_{\mu\nu}=\frac1{H^2u^2}\eta_{\mu\nu}\eqno(met a)$$
and
$$h_{\mu\nu}=\left(\frac1{Hu}\right)^{3-D/2}\chi_{\mu\nu}\eqno(met b)$$
It is not hard to show that the background metric $\hat g$ is
invariant under the de Sitter transformations \(xtrans). The perturbation
$h$ will transform as a tensor, so the rescaled quantity $\chi$ will
transform as
$$\delta\chi_{\mu\nu}=-\chi_{\mu\nu,\rho}\ \delta x^\rho
-(\frac D2-3)\frac1u\chi_{\mu\nu}\ \delta x^0-\chi_{\rho\nu}\ \delta x^
\rho_{,\mu}
-\chi_{\rho\mu}\ \delta x^\rho_{,\nu}\eqno(chitrans1)$$

We now wish to investigate the effect of these transformations on the gauge
fixed solutions to the equation of motion for $\chi$. Following Tsamis
and Woodard\refto{mode}, these solutions are
found by seperating $\chi$ into three parts, as
$$\chi_{ij}=\epsilon^{ij}_a+\frac1{D-3}\delta^{ij}(-\epsilon^{kk}_a+\epsilon_c
)\eqno(sepchi a)$$
$$\chi_{0i}=\epsilon_b^i\eqno(sepchi b)$$
$$\chi_{00}=\epsilon_c\eqno(sepchi c)$$
The gauge is fixed by setting $F_\mu=0$, where
$$F_\mu=\chi^\rho_{\mu,\rho}-\half\chi^\rho_{\rho,\mu}
-\left(\frac{D-2}{2u}\right)\chi^0_\mu-\left(\frac{D-2}{4u}\right)\delta_\mu^0
\ \chi^\rho_\rho\eqno()$$
This condition may be enforced by setting
$$\epsilon_c=\epsilon_b^i=\epsilon_a^{ii}=\epsilon_a^{ij,j}=0.\eqno(gauge)$$
This set of conditions is stronger than $F_\mu=0,$ but for solutions to the
linearized field equations, there is always enough
residual gauge invariance to impose \(gauge) after imposing $F_\mu=0$.
Then the equation of motion for $\epsilon_a$ is
$$D_a\epsilon_a\equiv\left(\partial^2+\frac{D(D-2)}{4u^2}\right)\epsilon_a=0
\eqno(epseq)$$
This equation is solved by
\def\ik{\int{d^{D-1}k\over(2\pi)^{D-1}}}
\def\cc{\hbox{ + c.c.}}
\def\vk{\vec k}
$$\epsilon_a^{ij}=\ik\left[A^{ij}(\vk)\ \chi_a(u,\vec x;\vk)\cc\right]
\eqno(epsexp)$$
where the basis functions $\chi_a$ have the form
$$\chi_a(u,\vec x,\vk)\equiv {\sqrt{\half\pi ku}}H^{(1)}_{{D-1\over 2}}(ku)
\ \exp(i\vk\cdot\vec x-i\frac kH+i\frac D4\pi)
\eqno(chiadef)$$% Chiadef, no relation to Chia Tze %

Since \(gauge) is clearly not de Sitter invariant, $\delta\chi$ as given by
\(chitrans1) will not obey the same gauge conditions as the original $\chi$.
It will be necessary to restore \(gauge) by gauge transformations. Thus the
full effect of the de Sitter transformation on a gauge fixed soultion will
be made up of three parts: $\delta_1$, given by \(chitrans1); a gauge
transformation $\delta_2$ to restore $F_\mu=0$; and another gauge
transformation
$\delta_3$ to restore \(gauge).

We now calculate explicitly the effect of \(chitrans1) on the $\epsilon$'s.
We find
$$\eqalignno{\delta_1\epsilon_a^{ij}=&-\frac{d^k}{2H}\chi^{ij,k}+c^{kl}(x^k
\chi^{ij,l}+\delta^{ik}\chi^{jl}+\delta^{jk}\chi^{il})\cr
&+b(x^k\chi^{ij,k}+u\chi^{ij,0}+(3-\frac D2)\chi^{ij})
+Ha^k\Bigl[-\half u^2\chi^{ij,k}+\half x^l x^l \chi^{ij,k}&(epstrans a)\cr
&-x^kx^l\chi^{ij,l}
-ux^k\chi^{ij,0}+(1-\frac D2)x^k\chi^{ij}+x^i\chi^{jk}
+x^j\chi^{ik}-x^l\delta^{ik}\chi^{jl}-x^l\delta^{jk}\chi^{il}\Bigr]}$$
$$\delta_1\epsilon_b^i=-Ha^j u\chi^{ij}\eqno(epstrans b)$$
$$\delta_1\epsilon_c=0\eqno(epstrans c)$$
Consequently, the change in the gauge fixing quantity $F_\mu$ is
$$\delta_1 F_0=0\eqno(Ftrans a)$$
$$\delta_1 F_i=2Ha^k\chi_{ik}\eqno(Frtans b)$$
We wish to find a gauge transformation $\delta_2$ which cancels these results.
Under a gauge transformation characterized by a parameter $e_\mu(u,\vec x)$,
the variation of $\chi$ is
$$\delta_2\ \chi_{\mu\nu}=-e_{\mu,\nu}-e_{\nu,\mu}
-\frac{D-2}{2u}(e_\mu\delta_\nu^0
+e_\nu\delta_\mu^0)+\frac2u\eta_{\mu\nu}e^0\eqno(chigaugetrans)$$
{}From \(chigaugetrans), we see that if we choose $e_0=0$, then $\chi_{00}=0$
will be preserved. Then we have
$$\delta_2\ \epsilon_c=0\eqno(epstrans2 a)$$
$$\delta_2\ \epsilon_b^i=-e_{i,0}-\frac{D-2}{2u}e_i\eqno(epstrans2 b)$$
$$\delta_2\ \epsilon_a^{ij}=-e_{i,j}-e_{j,i}+\delta_{ij}e_{k,k}
\eqno(epstrans2 c)$$
so
$$\delta_2F_0=0\eqno(Ftrans2 a)$$
$$\delta_2F_i=-D_a e_i\eqno(Ftrans2 b)$$
Therefore we need to look for an $e_i$ such that
$$D_ae_i=2Ha^k\chi_{ik}\eqno()$$
A solution is
$$e_i=Hu\alpha^j\ik{A^{ij}\over k^2}\left(\frac\partial{\partial u}+\frac{D-2}
{2u}\right)\chi_a\eqno(edef)$$
We can add any $e_i'$ such that $D_a e_i'=0$. Let us use this residual gauge
freedom to take $\epsilon_b$ to zero. Calculating what we have so far,
$$(\delta_1+\delta_2)\epsilon_b^i=-(D-1)H\alpha^j\ik{A^{ij}\over k^2}\chi_a
\eqno()$$
This can be easily cancelled by a gauge transformation $\delta_3$ parametrized
by
$$e_i'\equiv -(D-1)H\alpha^j\ik{A^{ij}\over k^2}\chi_a\eqno()$$

We have now completely restored the original gauge conditions. What remains is
the straightforward but tedious calculation of $\delta\epsilon_a$. We will
not give the details of this calculation, except to note that there will be
many factors of $\vec x$, which one must get rid of by writing
$$x_k\ \exp(i\vec k\cdot\vec x)=-i{\partial\over\partial k^k}\
\exp(i\vk\cdot\vec x)\eqno()$$
and integrating by parts on $k^k$. This of course generates derivatives of the
Hankel function $H^{(1)}(ku)$, but in the end these are all eliminated and one
is left with
$$\delta\epsilon_a^{ij}=\ik\ \delta A^{ij}\ \chi_a(u,\vec x;\vk)\eqno()$$
with
$$\delta A^{ij}=-\frac i{2H}k^kd^k A^{ij}+c^{kl}k^k{\partial A^{ij}\over
\partial k^l}+c^{ik}A^{jk}+c^{jk}A^{ik}+b\left(\frac{ik}H+2-\frac32D-k\cdot
{\partial\over\partial k}\right)A^{ij}$$
$$+Ha^k\left[-\frac{iD(D-2)k^k}{8k^2}+\frac{k^k}{Hk}-\frac{ik^k}{2H^2}
+\left(\frac{iD}2+\frac kH\right){\partial\over\partial k^k}
-\frac{ik^k}2{\partial^2\over\partial k^l\partial k^l}
+ik^l{\partial^2\over\partial k^k\partial k^l}\right]A^{ij}\eqno(hwat)$$
$$+Ha^k\left(\frac{iDk^i}{2k^2}+\frac{k^i}{Hk}
+i{\partial\over\partial k^i}\right)
A^{jk}-iHa^i{\partial A^{jl}\over\partial k^l}
+Ha^k\left(\frac{iDk^j}{2k^2}+\frac{k^j}{Hk}+i{\partial\over\partial
k^j}\right)
A^{ik}-iHa^j{\partial A^{il}\over\partial k^l}$$
It is straightforward to verify that this result satisfies the gauge conditions
$$\delta A^{ii}=k_j\delta A^{ij}=0\eqno()$$

In conclusion, we have shown that the set of solutions to the linearized gauge
fixed equations of motion are closed under de Sitter transformations, and have
given the explicit transformation laws for these solutions. It would be
interesting to classify these solutions into irreducible representations of the
de Sitter group. Equation
\(hwat) will be the starting point for any such analysis.

I thank R. P. Woodard for suggesting this problem, and for helpful discussions.
This work has been supported by the department of energy under contract
DOE-AS05-80ER-10713.

\references

\refis{structure}N. C. Tsamis and R. P. Woodard, ``The Structure of
Perturbative
Quantum Gravity on a De Sitter Background'', Florida preprint UFIFT-92-14, to
appear in {\sl Comm. Math. Phys.}

\refis{mode}N. C. Tsamis and R. P. Woodard, ``Mode Analysis and Ward Identities
for Perturbative Quantum Gravity in de Sitter Space'', Florida preprint
UFIFT-92-20, to appear in {\sl Phys. Lett. }{\bf B292,} Oct. 15 1992 issue.

\refis{strong}N. C. Tsamis and R. P. Woodard, ``Relaxing the Cosmological
Constant'', Florida preprint UFIFT-92-23.

\refis{weinberg}S. Weinberg, \rmp 61, 1989, 1.

\endreferences

\end